\documentclass[conference]{IEEEtran}
\IEEEoverridecommandlockouts

\usepackage{cite}
\usepackage{amsmath,amssymb,amsfonts}
\usepackage{algorithmic}
\usepackage{graphicx}
\usepackage{textcomp}
\usepackage{xcolor}

 \usepackage{multirow}
 \usepackage{makecell} 

\def\BibTeX{{\rm B\kern-.05em{\sc i\kern-.025em b}\kern-.08em
    T\kern-.1667em\lower.7ex\hbox{E}\kern-.125emX}}

\begin{document}

\title{Evaluating Structural and SAT Attack Resilience in Cryptographic Dynamic Logic Locking via Keystream Capping}

\author{\IEEEauthorblockN{Nahush Tambe, Naseeruddin Lodge, Vineet Chadalavada, Fareena Saqib}
\IEEEauthorblockA{\textit{Electrical and Computer Engineering, UNC Charlotte, Charlotte, USA} \\
Ntambe@charlotte.edu, Nlodge@charlotte.edu, Vchadala@charlotte.edu, Fsaqib@charlotte.edu}}

\maketitle

\begin{abstract}
Untrusted third-party entities throughout the semiconductor supply chain expose proprietary hardware intellectual property to reverse engineering, piracy, and overproduction. While logic locking counters this threat by embedding key-dependent logic, conventional static key-gates remain vulnerable to both oracle-guided SAT attacks and structural removal attacks due to their time-invariant netlist signatures. Dynamic schemes such as STREAMLOCK mitigate oracle attacks by continuously masking signals with cryptographic keystream, achieving near 50\% Hamming distance output corruption under incorrect keys with low hardware overhead. However, confining key-gates to primary outputs leaves internal combinational cones vulnerable to single-tap structural bypasses. This work extends dynamic logic locking by distributing key-gates within internal paths prioritized via fault-propagation analysis. The original and internally-hardened architectures are evaluated against structural removal attacks and a bounded, unrolling-based SAT-style attack across three ISCAS'85 benchmarks. Distributing masking internally raises the uninformed attacker's brute-force search space from 6 candidate node subsets to over $10^{45}$ for the largest benchmark. Crucially, empirical analysis reveals that topological relocation alone does not prevent algebraic key extraction when wide circuits expose numerous keystream bits per clock cycle. To eliminate this leak, a keystream-bit reuse cap is introduced to restrict per-cycle exposure independent of circuit width. Under this constraint, SAT-solver tractability encounters an immediate computational cliff, transitioning sharply from rapid under-constrained solving to timeout thresholds without disclosing the secret key. Integrating fault-aware internal placement with keystream reuse capping delivers joint resilience against structural bypasses and bounded SAT attacks.

\end{abstract}


\section{Introduction}
The globalization of semiconductor design and fabrication has made intellectual property (IP) protection a central concern in modern integrated circuit development \cite{Primer_Hardware_Security_paper}. Because physical fabrication relies on external third-party foundries and test facilities, hardware designs face significant exposure to reverse engineering, IP piracy, and unauthorized overproduction. Logic locking has emerged as a widely adopted countermeasure, embedding key-dependent gates into a netlist such that correct functionality is recovered only upon applying the valid activation key. However, most conventional schemes rely on static key-gates whose Boolean state does not evolve at runtime, leaving them susceptible to oracle-guided attacks that exploit the time-invariant relationship between key, input, and output. These schemes span early random-insertion methods such as Random Logic Locking (RLL) \cite{EPIC_paper}, as well as structured techniques such as SAT-Attack Resilient Logic Locking (SARLock) \cite{SARLock_paper}, fault-analysis-based logic encryption \cite{FALL_paper}, and Stripped Functionality Logic Locking (SFLL) \cite{From_Theory_To_Practice_paper}.

Motivated by this limitation, recent work has explored dynamic, cryptographically driven logic locking that replaces the static key-input-output relationship of conventional schemes with masking behavior that evolves continuously over time. STREAMLOCK \cite{STREAMLOCK_paper} introduces one such approach, generating a lightweight, 128-bit keystream from a nonlinear feedback shift register (NLFSR)-based stream cipher and using it to mask circuit outputs with a time-varying signal rather than a static key-gate value. This continuously evolving mask achieves strong functional corruption under incorrect keys while introducing minimal hardware overhead, effectively neutralizing the time-invariant assumptions that the conventional oracle attacks rely upon.  

Prior evaluations of dynamic, cryptographically driven schemes have centered primarily on output corruption under incorrect keys and resistance to oracle-guided key recovery, leaving their vulnerability to structural, netlist-level attacks largely unexamined. This work extends STREAMLOCK by distributing its key-dependent masking across internal, fault-propagation-selected nodes rather than concentrating it at the primary-output boundary. A structural removal analysis across ISCAS'85 benchmarks \cite{ISCAS85_paper} demonstrates that output-only masking leaves the underlying combinational logic directly accessible in the netlist, allowing an adversary to recover correct functionality without solving for the correct key. This exposes a structural vulnerability that cipher strength alone cannot compensate for, as the failure arises from where the mask is physically placed rather than how the keystream is generated. 

Distributing key-dependent logic throughout the circuit's internal structure mitigates this vulnerability by embedding dependency chains between masked nodes, preventing an adversary from isolating the protected logic cone through a single structural tap. However, an evaluation against both structural and oracle-guided attacks reveals that internal node relocation alone is insufficient because wide circuits continue to leak critical algebraic information to SAT solvers. To address this limitation, a keystream-bit reuse cap is introduced to bound per-cycle exposure independent of circuit width. By coupling fault-aware internal placement with keystream reuse capping, the proposed architecture achieves resilient, multi-vector protection while preserving the lightweight overhead essential to practical dynamic locking. 

\noindent The primary contributions of this work are summarized as follows:
\begin{itemize}
    \item \textbf{Structural Removal Analysis:} Demonstrates that output-only keystream masking exposes an uncorrupted combinational logic cone, allowing adversaries to bypass dynamic defenses via direct gate-level tapping without cryptanalysis.
    \item \textbf{Fault-Aware Distributed Hardening:} Integrates dynamic masking into internal circuit nets prioritized by stuck-at fault impact, establishing cascading functional interdependencies that neutralize localized, single-tap extraction.
    \item \textbf{Keystream-Bit Reuse Capping:} Analyzes the rapid convergence of unrolled SAT formulations on wide-output topologies and enforces a per-cycle bit-reuse ceiling that constrains algebraic information leakage independent of circuit dimensions.
    \item \textbf{Empirical Multi-Vector Attack Evaluation:} Assesses the original and hardened designs against physical netlist removal and bounded unrolling SAT solvers across three representative ISCAS'85 netlists, establishing the emergence of a sharp computational barrier under exposure constraints.
\end{itemize}

\section{Background \& Literature Review}

\subsection{Static Logic Locking \& Oracle-Guided Key Recovery}
Logic locking protects an integrated circuit by conditioning correct functional operation on the application of a valid secret key. Ideally, key-dependent gates are configured such that an incorrect key induces functional corruption across the input space rather than leaving outputs indistinguishable from golden behavior \cite{Evaluating_Security_Logic_Locking_paper}. Random-insertion techniques such as RLL \cite{EPIC_paper} insert key-gates arbitrarily without evaluating circuit depth or functional observability, yielding low output corruptibility and leaving candidate keys vulnerable to simple pruning algorithms. This vulnerability prompted structured architectures designed explicitly to resist Boolean satisfiability (SAT) attacks \cite{Evaluating_Security_Logic_Locking_paper}, which iteratively query an activated functional oracle for distinguishing input patterns (DIPs) to systematically rule out incorrect key hypotheses. Defenses such as SARLock \cite{SARLock_paper} and SFLL \cite{From_Theory_To_Practice_paper} neutralize SAT solvers by confining functional corruption to a sparse input subdomain, forcing an exponential number of DIP iterations at the cost of creating localized structural signatures. Functional analysis attacks \cite{Functional_Analysis_Attack_paper} exploit these isolated structural traces to extract secret keys directly without iterative oracle queries, demonstrating that point-function architectures remain vulnerable to netlist inspection.

Approximate and accelerated variants of the SAT attack address the iteration bottlenecks that arise in point-function schemes. AppSAT \cite{APPSAT_paper} relaxes the exactness requirement, terminating early once an approximate key falls below a target error rate. Similarly, Double-DIP \cite{Double_DIP_paper} eliminates at least two candidate keys per iteration to accelerate exact recovery against symmetric defenses. These relaxed formulations remain effective against point-function defenses because restricting functional corruption to a narrow input region allows an approximate key to exhibit near-zero error across standard operating conditions.

\subsection{Structural Attacks on Key-Dependent Logic}
A distinct class of attacks circumvents the oracle-guided model entirely by analyzing the gate-level netlist directly without querying an activated chip for distinguishing patterns. Removal attacks \cite{Removal_Attack_paper} identify topological or sensitization boundaries that isolate the key-dependent gates from the underlying combinational cone, allowing an adversary to excise or bypass the locking circuitry. In boundary-masked implementations, an adversary can tap intermediate nets immediately upstream of the output XOR gates, recovering golden functionality without extracting or guessing the secret key. Because this bypass requires no cryptanalytic effort, it exposes a physical failure mode where an architecture may withstand advanced SAT solvers yet remain trivially defeated due to topological separability.

A related yet mechanically distinct threat is the bypass attack \cite{Bypass_Attack_paper}, which primarily targets point-function architectures such as SARLock and SFLL. Rather than excising key-dependent logic, bypass attacks augment the netlist with corrective comparators that detect the sparse input patterns responsible for error injection, overriding faulty states during execution. Because dynamic masking perturbs outputs broadly across execution cycles rather than confining corruption to an isolated input subset, it eliminates the static signatures required by bypass comparators, leaving structural removal as the primary netlist-level threat evaluated in this work.

\subsection{STREAMLOCK: Stream Cipher-Driven Dynamic Locking}
To prevent adversaries from exploiting time-invariant input-output relationships, recent defenses replace static key-gates with time-varying cryptographic obfuscation. STREAMLOCK \cite{STREAMLOCK_paper} implements this mechanism by generating a continuously evolving, 128-bit keystream from four cross-coupled nonlinear feedback shift registers (NLFSRs) governed by a Geffe-inspired nonlinear combiner. The keystream generator is initialized with the secret activation key alongside a public initialization vector (IV) and advances every clock cycle, ensuring that the effective masking evolves continuously across successive operations. Dense cross-coupling among the four shift registers, combined with an initial warm-up interval that suppresses keystream emission until achieving full state diffusion, impedes algebraic decomposition and correlation attacks on the internal state \cite{Trivium_paper}.

This temporal variation fundamentally distinguishes STREAMLOCK from conventional static key-gate architectures. Standard oracle-guided SAT attacks assume that the locked circuit computes an invariant Boolean mapping across evaluation queries, whereas STREAMLOCK invalidates this premise by remapping the input-output relationship dynamically on every clock cycle. This time-varying design achieves strong functional corruption under incorrect keys, yielding approximately 50\% output Hamming distance relative to the golden circuit, while preserving full functional correctness under the correct key. Crucially, this corruption is realized with minimal hardware overhead compared to heavyweight alternatives such as AES-based locking \cite{NIST_AES, Heavyweight_AES_paper}. Because the keystream generator operates as a compact sequential primitive rather than an expansive combinational block \cite{Speed_Up_NLFSR_paper}, the architecture offers a practical balance between static key-gate vulnerabilities and prohibitive block-cipher silicon overhead.

The structural removal vulnerability analyzed in this work arises under an adversarial threat model distinct from the one originally targeted by the baseline framework. Removal attacks assume a white-box adversary with access to the gate-level netlist, capable of tracing signal hierarchies and isolating sub-circuits within an untrusted foundry or third-party facility. In contrast, oracle-guided adversaries operate in a black-box setting with access limited to input-output stimulus on a packaged chip, typical of untrusted end users or downstream integrators. Against this oracle-only threat model, the original output-boundary design remains fully effective because its time-varying keystream defeats the fixed-Boolean-function assumption underlying SAT-based key recovery regardless of where masking is applied. The internal, fault-aware distribution introduced in this work is therefore best understood as extending protection to a stronger adversary with gate-level visibility, rather than correcting an inherent flaw under black-box assumptions.

In its baseline configuration, the generated keystream masks the circuit exclusively at its primary outputs, preserving internal timing across the combinational core while defeating oracle-guided adversaries. However, against a netlist-level adversary, leaving the upstream combinational core physically unaltered exposes a distinct boundary that allows structural removal attacks to bypass the defense, irrespective of cipher strength.

\section{Proposed Methodology}

The proposed fault-aware extension of STREAMLOCK distributes key-dependent masking across internal high-impact nodes selected via fault-propagation analysis, rather than concentrating it at the primary-output boundary. In addition, the architecture bounds per-cycle keystream exposure by enforcing a keystream-bit reuse cap that operates independently of circuit width. The resulting framework is evaluated against a re-scoped removal attack that quantifies escalating attacker search cost rather than claiming absolute prevention. Furthermore, it is evaluated against a bounded, unrolling-based SAT-style attack that directly tests the time-varying keystream against the static-Boolean assumptions of conventional solvers. The remainder of this section details the fault-aware node-selection algorithm, the mechanics of distributed key-gate insertion, the keystream-bit reuse capping formulation, and the dual evaluation methodologies for structural removal and bounded SAT unrolling.

\subsection{Node Selection via Fault-Propagation Analysis}
Fault-propagation analysis guides the identification of internal key-gate locations. For every candidate internal net, excluding primary inputs and outputs, stuck-at-0 and stuck-at-1 faults are injected, and the resulting primary-output corruption relative to the golden circuit is measured across a shared test-vector set. A dedicated fault simulator based on the stuck-at evaluation methodology of HOPE \cite{HOPE_paper} conducts this characterization. Candidate nets are subsequently ranked by their cumulative output corruption impact, with the top $N$ nets selected for key-gate insertion. Constraining $N$ to match each benchmark's baseline primary-output key count isolates internal relocation as the sole variable governing any observed security enhancement.

\subsection{Distributed Key-Gate Insertion}
For each selected net $X$, the keystream masking stage is relocated from the circuit boundary directly to $X$. The driver of net $X$ is rerouted to an intermediate net $X_{\text{pre}}$, and a two-input XOR gate implementing $X = X_{\text{pre}} \oplus z_i(t)$ is inserted, where $z_i(t)$ represents the $i$-th bit of the dynamic 128-bit keystream word generated at cycle $t$. Because all downstream gates continue to reference $X$ as their input, any functional corruption injected at $X$ propagates naturally throughout its entire fanout cone. When a masked net drives downstream logic cones that contain additional key-gates, compounding dependency chains emerge across the netlist. This structural coupling marks a fundamental architectural departure from boundary masking, where key-gates operate as isolated terminal stages devoid of inter-gate feedback or logical depth.

\subsection{Keystream-Bit Reuse Capping}
Relocating key-dependent masking to internal nodes alters the topological position of the key-gates within the netlist, but does not inherently reduce the volume of distinct keystream bits exposed per cycle. Without reuse constraints, internal masking applies up to one unique keystream bit per insertion site during every operational cycle, matching the per-cycle exposure rate of boundary masking. Because bounded SAT-style solvers exploit this per-cycle observation bandwidth to constrain the underlying stream generator, an unconstrained distributed architecture remains vulnerable to key recovery across extended time unrollings.

Mitigating this algebraic exposure requires bounding the per-cycle keystream bandwidth directly. Under the bit-reuse capping scheme, the $i$-th masked location is driven by keystream bit $z_{i \bmod K}(t)$ rather than an unconstrained $z_i(t)$, where $K$ denotes the maximum allowable distinct keystream bits deployed in any single clock cycle. In this work, $K$ is configured to $7$ as a security threshold chosen to suppress solver tractability across extended unrolling depths. For circuits where the total count of masked locations does not exceed $K$, including benchmarks C17 and C432, every insertion location retains a unique keystream bit. For wider circuits, multiple internal nodes share keystream bits within each execution cycle, effectively bounding the rate of fresh algebraic information exposed to an adversary independently of circuit width.

\subsection{Removal-Attack Re-Evaluation}
Following the distribution of key-dependent masking across internal nodes, the removal-attack methodology \cite{Removal_Attack_paper} is re-evaluated against the modified architecture. Because internal-node masking relies on combinational XOR operations, an attacker with complete knowledge of the selected node set can, in principle, reconstruct the golden function by tracing and correcting each net individually. Two distinct adversarial capabilities are evaluated to frame defensive strength around escalating search complexity rather than an unrealistic claim of absolute prevention. The informed adversary represents a theoretical worst-case bound with prior knowledge of the targeted nets, whereas the uninformed adversary models a realistic scenario where an attacker must discover the embedded insertion sites across the entire combinational fabric.

In the baseline design, the original combinational cones feed directly into a boundary keystream XOR stage immediately preceding the primary outputs. Because the net immediately upstream of each XOR gate retains golden functionality, an adversary possessing netlist visibility can tap these wires directly, bypassing the locking mechanism without recovering the secret key. The locked outputs, by contrast, mismatch the golden circuit across the majority of test patterns, confirming that dynamic masking operates effectively against any observer that does not structurally isolate the underlying logic. This direct tapping vulnerability motivates the internal, dependency-driven distribution implemented in the hardened architecture.

\subsubsection{Informed-Attacker: Naive vs. Corrected Reconstruction}
This scenario evaluates whether the baseline removal attack, applying an isolated hierarchical tap per insertion point, still succeeds once masking moves away from the primary-output boundary. Security against this adversary is quantified by evaluating two distinct reconstruction models against the golden circuit. A \textit{naive} reconstruction taps the pre-mask intermediate net $X_{i,pre}$ directly at each insertion point, mirroring the simple tapping strategy of the original attack. A \textit{corrected} reconstruction accounts for topological interactions, recomputing every intermediate net located downstream of another masked location with the corrected upstream value before determining the final primary outputs. This comparison isolates instances where masked nodes are mutually independent from cases where cascading internal dependency chains require recursive structural re-synthesis.

Consider a representative case where masking is distributed across two dependent nets, $X_1$ and $X_2$, with $X_2$ positioned downstream of $X_1$ along a shared signal path. An adversary who taps both pre-mask nets $X_{1,pre}$ and $X_{2,pre}$ under the assumption of mutual independence reconstructs $X_1$ correctly, yet derives $X_{2,pre}$ from logic that consumes the corrupted, runtime-masked value of $X_1$. This leaves the primary outputs corrupted despite obtaining both physical taps. Golden circuit functionality is recovered only by explicitly re-evaluating the intermediate logic feeding $X_2$ using the corrected value of $X_1$, resolving the topological dependency chain. This behavior confirms that internal masking cannot be bypassed simply by scaling the attacker's tap count, as defeating the scheme requires an additional recursive re-synthesis step that has no counterpart in boundary masking.

\subsubsection{Uninformed-Attacker and Search Space Cost}
This scenario models an adversary who lacks prior knowledge of the targeted node set and must identify the modified locations directly from an extracted, flattened gate-level netlist. Identifying the exact locked configuration requires searching across $\binom{M}{N}$ candidate subsets, where $M$ denotes the total number of internal candidate nets and $N$ represents the number of inserted key-gates. Each candidate combination demands independent functional verification against observed input-output responses to confirm circuit restoration. In contrast, the search space for boundary masking collapses to a single trivial configuration, as the masked signals are directly accessible at the primary-output interface. While this combinatorial bound primarily governs adversaries unable to replicate the proprietary fault-simulation ranking, it establishes a formal lower bound on brute-force netlist deobfuscation complexity.

\subsection{Bounded SAT-Style Attack Formulation}
The central premise of this architecture is that a time-varying keystream invalidates the static Boolean function assumption underlying conventional SAT-based key recovery \cite{Evaluating_Security_Logic_Locking_paper}. To ground this claim empirically, a bounded attack framework is formulated to evaluate both the baseline output-only design and the reuse-capped architecture, isolating the direct impact of bounded keystream exposure on solver tractability. The adversary is assumed to possess gate-level netlist visibility, exact knowledge of the keystream mapping at each insertion site, and access to the golden intermediate states extracted via structural isolation. The 128-bit secret key remains unknown, while the initialization vector (IV) is treated as public. The adversary samples $D$ consecutive operational cycles while the keystream generator advances on every clock edge. The exposed keystream bits are reconstructed via known-plaintext XOR against observed node values. These extracted bits are encoded into a bounded Satisfiability Modulo Theories (SMT) constraint model incorporating the exact nonlinear feedback shift register update functions and Geffe-combiner over 32-bit bitvectors. The formulation unrolls the generator across $D$ consecutive clock cycles, treating the 128-bit secret key as the sole symbolic variable.

This symbolic formulation is constructed within an execution framework that guarantees direct equivalence with hardware behavior. Each functional primitive evaluates concrete bitwise operations or Z3 symbolic bitvectors, enforcing identical constraint semantics between the functional hardware specification and the solver encoding. The concrete execution path is validated bit-for-bit against cycle-accurate RTL simulations across extended operational sequences, while the symbolic path is verified to yield matching outputs when evaluated with the golden key. Furthermore, every unrolled formula is verified for satisfiability under the correct key to guarantee that the generated constraints do not overconstrain the system or inadvertently eliminate the true solution.

The unroll depth $D$ is swept systematically, dispatching each formulation to the Z3 solver under a fixed execution timeout across multiple combinational benchmark circuits of varying scale. Because these evaluated netlists incorporate distinct masked node counts ranging up to 22 internal sites, sweeping $D$ isolates how solver tractability scales with per-cycle leakage and quantifies the protective efficacy of capping runtime bit exposure at $K=7$.

\section{Results \& Discussion}
To evaluate the resilience of the fault-aware distributed architecture, an empirical removal attack first demonstrates the structural vulnerability of output-only masking, establishing experimental motivation for internally distributing key-gates. Functional and security verifications of the resulting architecture are then conducted by re-assessing structural removal complexity and deploying bounded SAT-style key-recovery attacks across ISCAS'85 benchmark circuits of varying scale. Key-dependent masking is confirmed to maintain functional corruption after relocating the gates from the primary outputs to internal nodes, and the resulting escalation in attack complexity is quantified under informed and uninformed adversary models. The bounded SAT-style attack is evaluated against the distributed architecture without keystream-bit reuse capping, demonstrating that internal relocation alone does not guarantee resistance for wider circuits. Finally, repeating the attack with bit-reuse capping applied demonstrates bounded algebraic leakage and characterizes solver-time scaling under time-varying keystream execution.

\subsection{Empirical Vulnerability of Output-Only Masking}
The structural exposure of output-only dynamic masking is evaluated directly through an empirical removal attack across three representative ISCAS'85 benchmarks, namely C17, C432, and C3540. For each circuit, a golden reference module and the locked module operate synchronously under a shared clock and common primary-input stimuli. The locked module preserves the original combinational core as an internal sub-circuit that directly feeds the terminal keystream XOR stage. Because this combinational logic remains unmodified during synthesis, its pre-mask signals remain fully accessible within the flattened netlist hierarchy. The removal attack exploits this topological boundary by referencing those pre-mask nets directly, completely bypassing the keystream XOR stage without querying an oracle or resolving the 128-bit key.

Both design instances are driven with identical input sequences, spanning an exhaustive 32-vector set for C17 and 200 pseudo-random vectors for C432 and C3540. These stimuli are applied following an initial warm-up phase of the keystream generator to ensure complete internal state diffusion. The golden outputs are compared cycle by cycle against both the tapped intermediate nets and the final masked outputs. To verify that structural extraction operates completely independent of key correctness, C432 is evaluated with an intentionally invalid key and initialization vector. Table \ref{tab:STREAMLOCK_removal_attack} summarizes these comparative results.

\begin{table}[h]
\centering
\caption{\centering Structural Removal Attack on Output-Only STREAMLOCK}
\scriptsize
\begin{tabular}{|c|c|c|c|}
\hline
\textbf{Circuit} & \textbf{Vectors Tested} & \makecell{\textbf{Bypassed-Path}\\\textbf{Mismatches}} & \makecell{\textbf{Locked-Output}\\\textbf{Mismatches}} \\
\hline
C17    & 32 (exhaustive) & 0 / 32   & 27 / 32  \\
C432   & 200 (random)*    & 0 / 200  & 199 / 200 \\
C3540  & 200 (random)    & 0 / 200  & 200 / 200 \\
\hline
\end{tabular}
\vspace{2pt}
\begin{center}
\parbox{0.9\columnwidth}{\centering\scriptsize *Evaluated with an incorrect key and IV supplied at the top level.}
\end{center}
\label{tab:STREAMLOCK_removal_attack}
\end{table}

Across all three circuits, the bypassed signal path reproduces golden circuit functionality with zero mismatches, confirming that the unmasked logic cone remains fully intact and directly recoverable once an adversary isolates it within the netlist. The locked outputs, by contrast, exhibit corruptive mismatches across the vast majority of tested vectors regardless of circuit size, demonstrating that dynamic masking operates effectively against any observer that does not structurally circumvent the mechanism. This empirical outcome motivates distributing dynamic masking beyond the terminal output boundary and into internal, high-impact logic paths.

\begin{figure}[h]
\centering
\includegraphics[width=0.85\columnwidth]{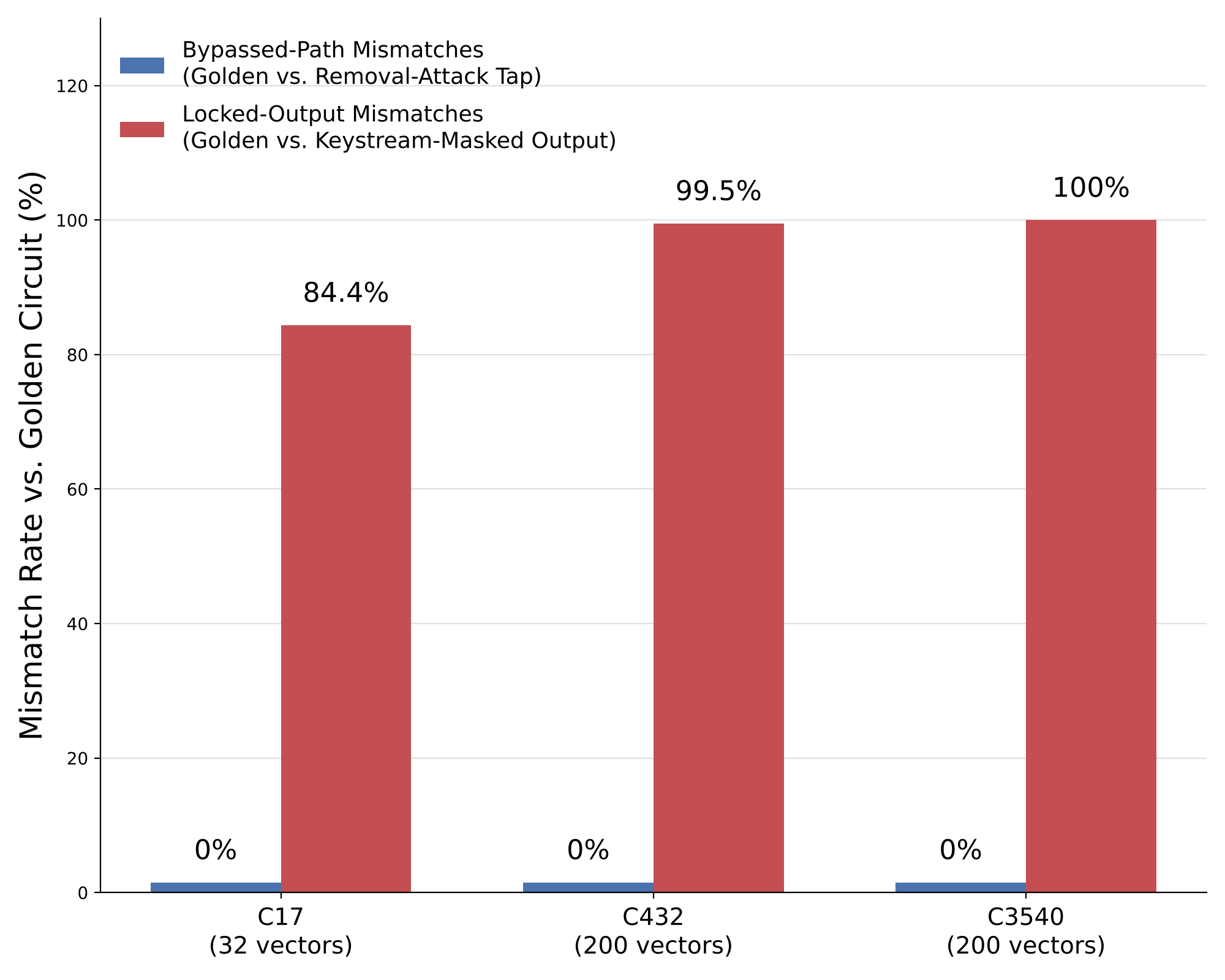}
\caption{\centering Structural Removal Attack Evaluation: Output-Only Masking}
\label{fig:removal_attack_plot}
\end{figure}

Figure \ref{fig:removal_attack_plot} illustrates this performance contrast across all three circuits.  The bypassed signal path reproduces golden circuit behavior exactly, maintaining a 0\% mismatch rate regardless of circuit size. In contrast, the locked-output mismatch rate reaches 84.4\% for C17 and approaches 100\% for C432 and C3540. This divergence confirms that dynamic masking functions effectively, and that the observed vulnerability stems entirely from boundary placement rather than cryptographic weakness.

\subsection{Removal-Attack Re-Evaluation on Distributed Architecture}

\begin{figure}[h]
\centering
\includegraphics[width=0.85\columnwidth]{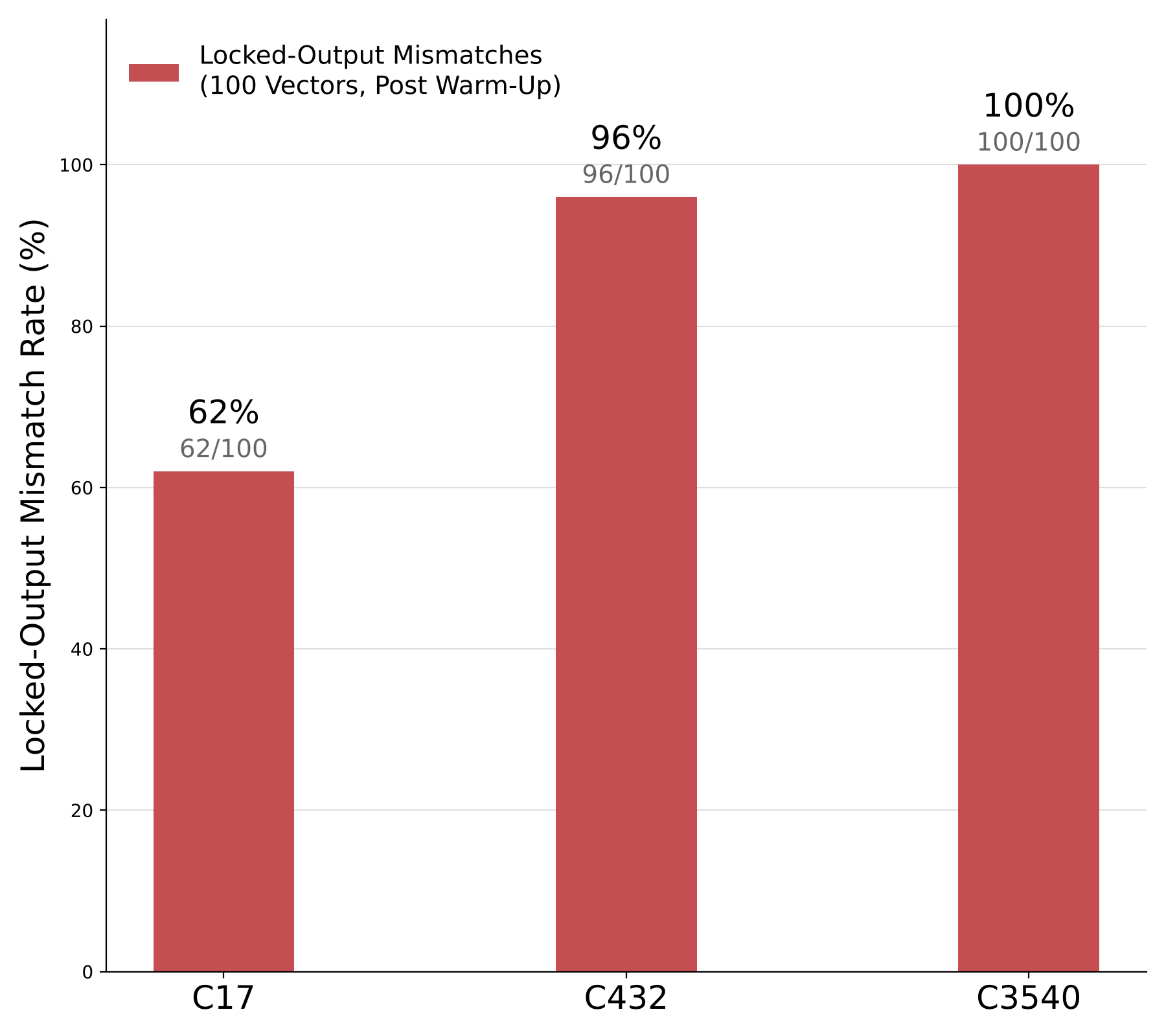}
\caption{Locked-Output Mismatch Rate for the Hardened Architecture}
\label{fig:informed_mismatch}
\end{figure}

Before evaluating attack complexity, empirical verification confirms that relocating key-gates from the primary-output boundary to internal nodes preserves functional corruptibility across downstream logic cones. Figure \ref{fig:informed_mismatch} reports locked-output mismatch rates measured against the golden circuit across 100 pseudo-random vectors following the keystream generator warm-up sequence.

Mismatch rates remain substantial across all evaluated circuits, confirming that relocating key-gates preserves functional corruptibility throughout the combinational core. Because internal masking corrupts intermediate nets that propagate through downstream logic cones, output corruption rates scale with circuit depth and structural complexity rather than remaining isolated at the interface. While reconvergent fault masking can occasionally produce a fortuitous golden output, the probability of such masking diminishes rapidly as the number of affected logic paths and primary outputs expands, driving the observed mismatch rate toward 100\% in larger benchmarks.

Under the dual-adversary framework, Table \ref{tab:naive_vs_corrected} reports the informed-attacker reconstruction results across all three benchmark circuits. Full golden circuit recovery remains achievable across all three benchmarks once the adversary systematically resolves the topological dependencies between inserted key-gates. Under the naive extraction approach, treating internal taps as mutually independent yields substantial corruption rates of 13 out of 32 for C17, 83 out of 100 for C432, and 99 out of 100 for C3540. The escalation in naive mismatch rates illustrates how downstream logic cones compound intermediate corruptions whenever upstream key-gates remain uncorrected. Consequently, the original attack methodology of applying isolated, boundary-level taps fails completely against the distributed architecture. To achieve zero functional mismatches, an adversary must recursively analyze the netlist to map cascading dependency chains, transforming a trivial single-step bypass into a complex structural deobfuscation workflow.

\begin{table}[h]
\centering
\caption{Informed-Attacker Evaluation Comparing Naive and Corrected Reconstructions}
\label{tab:naive_vs_corrected}
\scriptsize
\begin{tabular}{|c|c|c|}
\hline
\textbf{Circuit} & \makecell{\textbf{Naive Single-Tap}\\\textbf{Mismatches}} & \makecell{\textbf{Corrected Reconstruction}\\\textbf{Mismatches}} \\
\hline
C17   & 13 / 32  & 0 / 32  \\
C432  & 83 / 100 & 0 / 100 \\
C3540 & 99 / 100 & 0 / 100 \\
\hline
\end{tabular}
\end{table}

Full golden circuit recovery remains achievable across all three benchmarks once the adversary systematically resolves the topological dependencies between inserted key-gates. Under the naive extraction approach, treating internal taps as mutually independent yields severe corruption rates of 13 out of 32 for C17, 83 out of 100 for C432, and 99 out of 100 for C3540. The escalation in naive mismatch rates illustrates how downstream logic cones compound intermediate corruptions whenever upstream key-gates remain uncorrected. Consequently, the original attack methodology of applying isolated, boundary-level taps fails completely against the distributed architecture. To achieve zero functional mismatches, an adversary must recursively analyze the netlist to map cascading dependency chains, transforming a trivial single-step bypass into a complex structural de-obfuscation workflow.

\begin{table}[h]
\centering
\caption{Uninformed-Attacker Combinatorial Search Space Across Benchmarks}
\label{tab:search_space}
\scriptsize
\begin{tabular}{|c|c|}
\hline
\textbf{Circuit Configuration} & \textbf{Candidate Subsets $\binom{M}{N}$} \\
\hline
C17 (output-only baseline) & 1 (self-evident) \\
C17 (distributed)          & 6 \\
C432 (distributed)         & $3.39 \times 10^{11}$ \\
C3540 (distributed)        & $1.12 \times 10^{45}$ \\
\hline
\end{tabular}
\end{table}

\begin{figure}[h]
\centering
\includegraphics[width=0.85\columnwidth]{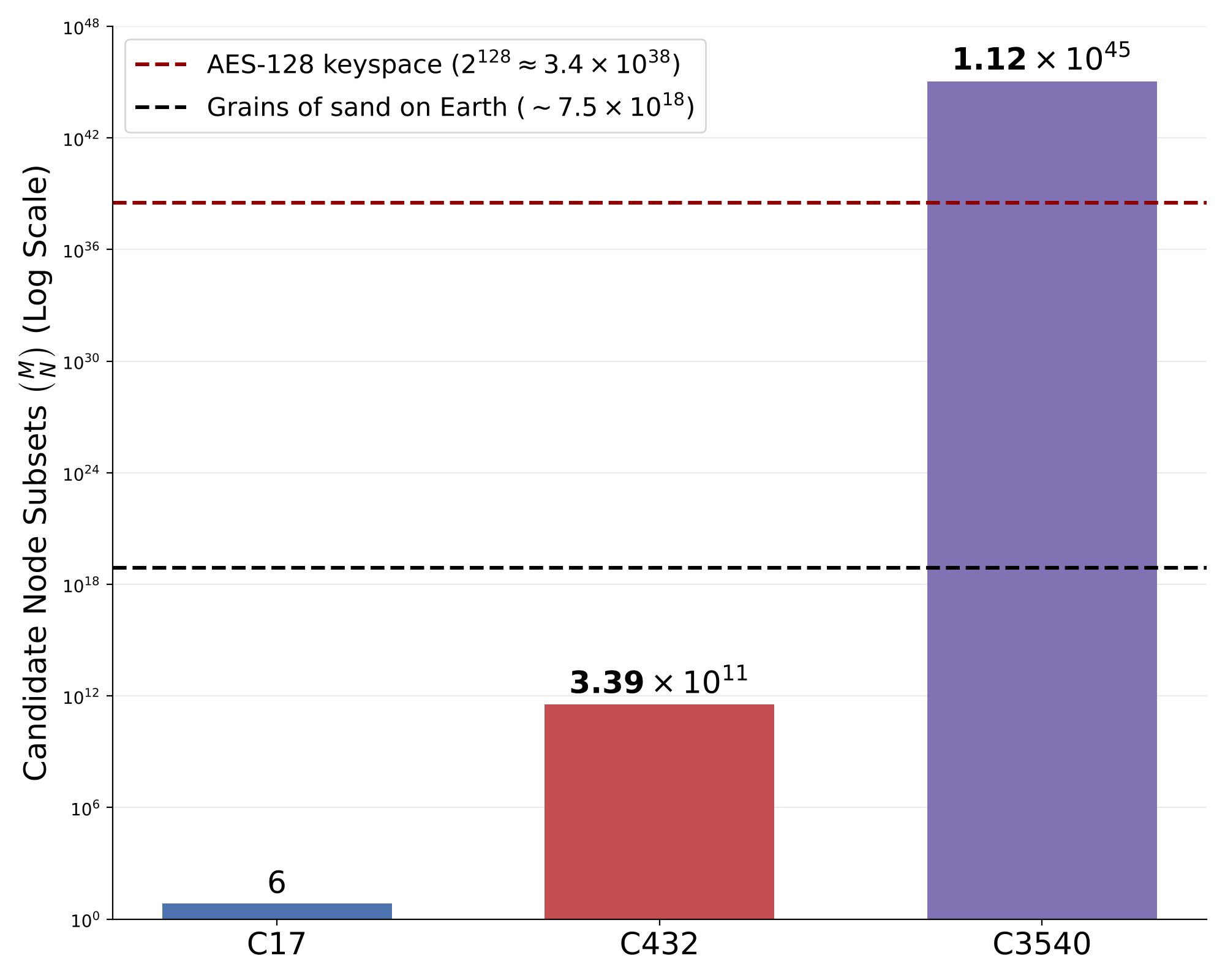}
\caption{Uninformed-Attacker Search-Space Growth Across Circuits}
\label{fig:search_space}
\end{figure}

Table \ref{tab:search_space} reports the uninformed-attacker search-space size across all three benchmark circuits, while Figure \ref{fig:search_space} illustrates the combinatorial scaling. The search space expands sharply with circuit scale. For C17, the six candidate subsets remain trivial enough for an uninformed attacker to evaluate exhaustively in seconds, offering only modest improvement over the self-evident baseline. Conversely, C432 and C3540 exceed the reach of brute-force enumeration, with C3540 pushing candidate combinations past the AES-128 keyspace. This progression reflects the rapid increase in internal candidate nets $M$ relative to the comparatively small key-gate count $N$ embedded across the netlist. Consequently, the uninformed-attacker defense strengthens considerably as designs scale upward, despite the limited margin demonstrated by elementary benchmarks like C17. These combinatorial figures characterize structural scaling trends with circuit dimensions rather than formal cryptographic entropy, and are not presented as equivalent to cryptographic security parameters.

Importantly, this combinatorial threshold holds solely against adversaries who do not reproduce the fault-propagation ranking algorithm. Under a strict Kerckhoffs-compliant threat model where the node-selection methodology is fully disclosed, an adversary who re-runs identical fault simulations against the extracted netlist isolates the targeted internal nodes with minimal computational overhead. The combinatorial growth is therefore framed as an operational barrier against uninformed reverse-engineering attempts rather than an absolute guarantee of structural concealment.

\subsection{SAT-Style Attack Evaluation on Distributed Architecture}

Applied directly to the distributed architecture without keystream-bit reuse capping, the bounded SAT-style attack recovers the true 128-bit key on C3540 at unroll depth 8 in 1.3 seconds across repeated executions, where 22 masked locations are evaluated per cycle. Relocating key-dependent masking to internal nodes does not independently preserve the key-recovery resistance observed on narrower benchmarks such as C17 and C432. Solver difficulty is governed primarily by the volume of unique keystream bits exposed per cycle rather than the topological location where the XOR operations occur. Consequently, wide combinational circuits that expose substantial algebraic constraints within a few operational cycles remain vulnerable to SMT unrolling regardless of netlist distribution depth. This empirical vulnerability directly motivates the keystream-bit reuse capping mechanism.

Table \ref{tab:sat_attack} reports solver execution characteristics when keystream-bit reuse capping is applied, restricting per-cycle exposure to at most $K=7$ distinct keystream bits regardless of native masked-location count. The primary objective is evaluating whether computational difficulty scales gradually with unrolling depth or exhibits an abrupt threshold transition characteristic of time-varying cryptographic primitives.

\begin{table}[h]
\centering
\caption{\centering SAT Attack Solver Behavior vs. Unroll Depth under Reuse-Capped Architecture}
\scriptsize
\begin{tabular}{|c|c|c|c|}
\hline
\textbf{Circuit} & \makecell{\textbf{Capped Bits}\\\textbf{per Cycle}} & \makecell{\textbf{Depths Solved}\\\textbf{(SAT, non-matching)}} & \makecell{\textbf{First TIMEOUT}\\\textbf{Depth ($\geq$60s)}} \\
\hline
C17    & 2 & 1, 2, 4, 8            & 16 \\
C432   & 7 & 1, 2, 4, 6, 8         & 12 \\
C3540  & 7 & 1, 2, 3, 4, 6, 8, 10  & 12 \\
\hline
\end{tabular}
\vspace{2pt}
\begin{center}
\parbox{0.95\columnwidth}{\centering\scriptsize *Native masked-location counts before capping are 2 for C17, 7 for C432, and 22 for C3540.}
\end{center}
\label{tab:sat_attack}
\end{table}

\begin{figure}[h]
\centering
\includegraphics[width=0.85\columnwidth]{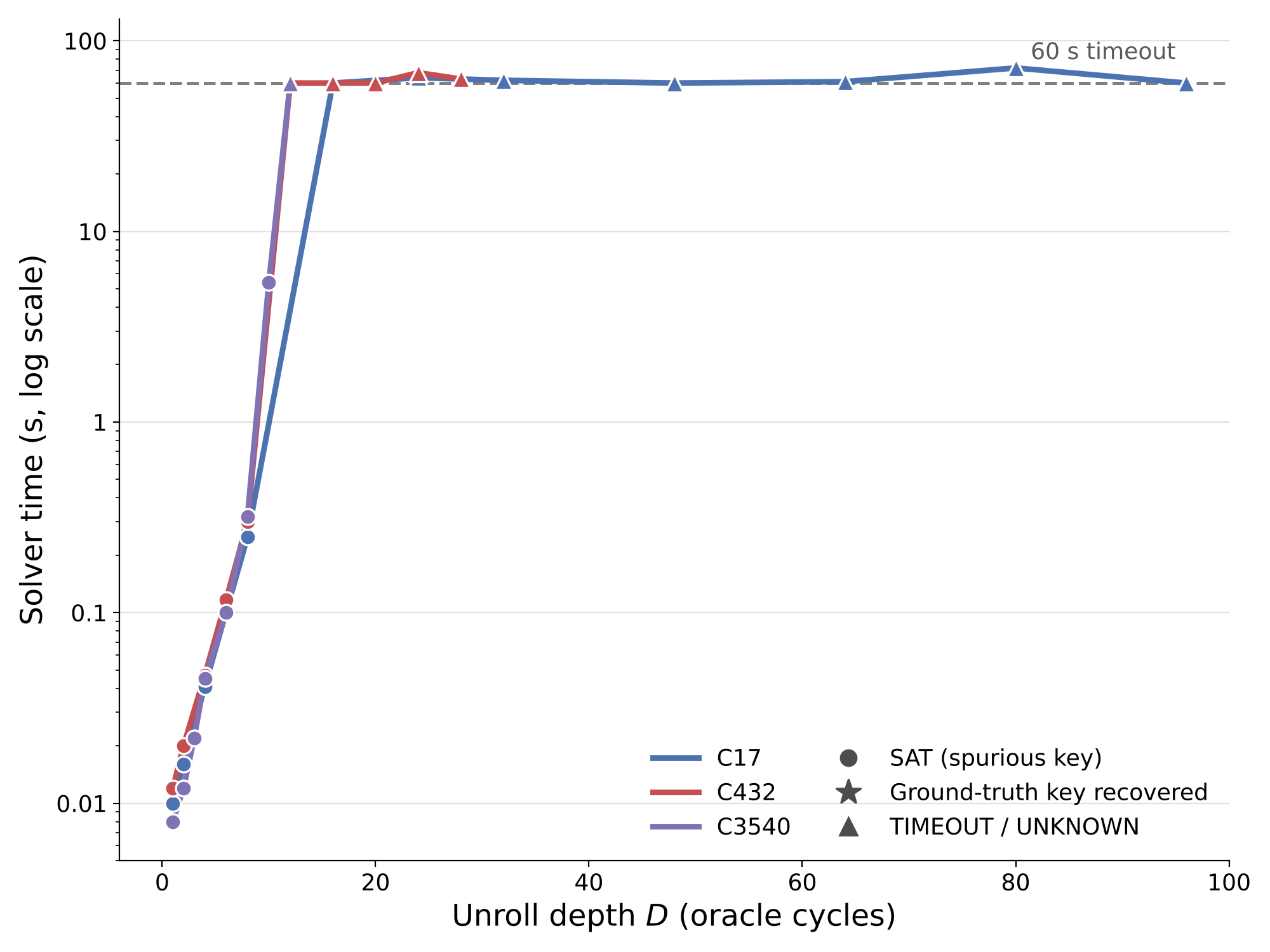}
\caption{Solve Time vs. Unroll Depth for the Bounded SAT-Style Attack}
\label{fig:sat_attack}
\end{figure}

Across all evaluated designs, from C17 with 2 native masked locations to C3540 with 22, the solver exhibits an identical sharp computational cliff. Under-constrained formulations yield rapid satisfiability results with non-matching, spurious keys through unroll depths 8 to 10, followed by an immediate transition to the 60-second timeout ceiling by depths 12 to 16. No capped configuration yields the true golden key or converges to an isolated unique key within the bounded evaluation window (Table \ref{tab:sat_attack}, Figure \ref{fig:sat_attack}). Most notably, C3540, which was broken outright at depth 8 when uncapped, exhibits the exact same timeout boundary at depth 12 as C432 under capping. Because C432 natively exposes 7 bits per cycle, this convergence demonstrates that SAT-attack tractability is governed strictly by the per-cycle keystream exposure cap rather than overall combinational circuit size once the two variables are decoupled.

Keystream capping directly accounts for the vulnerability observed in the uncapped evaluation, where netlists masking more locations per cycle yielded rapid key extraction at shallower unrolling depths rather than deeper ones. By constraining wide combinational circuits to an identical per-cycle exposure rate, the timeout depth converges consistently. This behavior confirms that the earlier divergence was governed entirely by the volume of algebraically exposed keystream bits per cycle rather than any intrinsic characteristic of circuit dimension or combinational topology.

\section{Conclusion}

This work extends STREAMLOCK dynamic logic locking by embedding key-dependent masking into internal circuit paths rather than restricting it to the primary-output boundary. Empirical evaluation demonstrates that output-only masking leaves the combinational logic cone structurally intact, permitting unmasked extraction through a single hierarchical tap regardless of key knowledge. Distributing key-gates across internal nodes identified through fault-propagation analysis eliminates this direct tap point, expanding the uninformed-adversary candidate search space from six combinations on small benchmarks to $1.12 \times 10^{45}$ on larger netlists. As emphasized throughout the evaluation, this combinatorial growth characterizes structural scaling trends rather than formal cryptographic entropy.

Internal node distribution alone does not mitigate bounded oracle-guided key recovery. Without reuse constraints, the widest evaluated circuit exposed sufficient algebraic relations per cycle to permit outright recovery of the 128-bit key under SMT unrolling, demonstrating that solver tractability depends on per-cycle keystream exposure rather than physical gate placement. Capping this exposure rate to seven distinct keystream bits per cycle resolves this vulnerability across wide combinational designs. Under reuse capping, SAT-solver execution across all three benchmarks exhibits rapid under-constrained solving followed by a sharp transition to the 60-second timeout ceiling, with no configuration yielding the ground-truth key. This abrupt threshold reflects the computational barrier inherent to time-varying dynamic locking primitives. Integrating fault-aware internal key-gate placement with keystream-bit reuse capping provides measurable resistance against both structural removal attacks and bounded SAT formulations while preserving dynamic execution properties.

\bibliographystyle{IEEEtran}
\bibliography{refs}  

\end{document}